# Strain Effects on the Mechanical Properties of Group-V Monolayers with Buckled Honeycomb Structures

Gang Liu,[a*] Zhibin Gao,[b*] Jian Zhou[c]

[a] School of Physics and Engineering, Henan University of Science and Technology, Luoyang 471023, China

[b] Department of Physics, National University of Singapore, Singapore 117551, Republic of Singapore

[c] National Laboratory of Solid State Microstructures and Department of Materials Science and Engineering, Nanjing University, Nanjing 210093, People's Republic of China

(*) Corresponding Author: liugang8105@gmail.com; zhibin.gao@nus.edu.sg

## Abstract

Based on first-principles calculations, we study systematically the ideal tensile stress-strain relations of three monoatomic group-V monolayer two dimensional (2D) materials with buckled honeycomb lattices: blue phosphorene, arsenene, and antimonene. The ideal strengths and critical strains for these 2D materials are investigated under uniaxial and equibiaxial strains. It is found that the ideal strengths decrease significantly as the atomic number increases, while the critical strains change not so much. In particular, the ideal strength of antimonene along armchair direction is found to exceed Griffith strength limit. The distributions of charge density, buckling heights, bond lengths and bond angles are also studied under different types of strains. It can be concluded that the critical strain is determined by the stretch and rotation of bonds simultaneously. Furthermore, the phonon dispersions, phonon instabilities, and failure mechanism of these materials under three types of strains are also calculated and explored.

## Introduction

2D material has been one of the most active areas of nanomaterials research since the discovery of graphene in 2004.[1] Due to numerous striking physical properties related to the low dimensionality of the quantum confinement effect, they have wide potential applications in next-generation electronic and energy conversion devices.[2-4] Furthermore, unexpected mechanical, such as negative Poisson's ratio and ultralow thermal conductivity in novel 2D materials also attract much attention.[5,6] Now the family of 2D materials has undergone rapid expansion. Many 2D materials have been synthesized experimentally and investigated in depth, such as hexagonal boron nitride,[7] transition metal dichalcogenides (TMDs),[8-11] group-IV monolayer materials,[12-15] and group-V monolayer materials.[16-25] It is believed there are two stable structures, buckled and puckered honeycomb structures for 2D group-V materials. Black phosphorene with puckered honeycomb structure is found to be a semiconductor with large direct band gap and high carrier mobility,[16,23,24] making it promising for potential applications.[25] Other 2D group-V materials with buckled honeycomb structures, such as blue phosphorene, arsenene and antimonene, have also drawn research interest due to their numerous novel physical properties.[16-22] To be specific, blue phosphorene, arsenene, and antimonene have flexible band gaps even higher than 2 eV, which are dependent on the number of layers and can be easily tuned by applying strain.[16-18,20,22] Furthermore, these materials can be transformed from indirect into direct band-gap semiconductors under small biaxial strain.[17,20] These unique properties make these buckled 2D group-V materials good candidates for the next generation of nanoelectronic and optoelectronic devices.

Generally, the practical application of any 2D material for designing and manufacturing devices hinges on a better understanding of its strength and mechanical behavior.[26,27] Ideal strength is the maximum stress that a perfect crystal can withstand at zero temperature and provides a measure for the intrinsic strength of the chemical bonding and overall stability of a material.[28-30] Furthermore, the band gap of these 2D group-V materials can be tuned by strain, as previously mentioned. Therefore, it is important to understand their elastic limit and how structures vary with strain. Unfortunately, most studies on 2D group-V materials have focused on their electronic properties.

In this work, we presented systematic analysis on the strain-induced mechanical properties of buckled blue phosphorene, arsenene, and antimonene, including the ideal strengths and corresponding critical strains, the variations of buckling heights, bond lengths and bond angles. The distributions of charge density are also investigated. Compared with allotropes, it can be argued the critical strains are affected by structures significantly. The ideal strength of antimonene along armchair (AC) direction exceeds Griffith strength limit.[31] It can be concluded that critical strain is determined by the stretch and rotation of bonds simultaneously. Under a certain type of critical strain, the more the bond is stretched, the less it rotates, and vice versa. It is also found the bond lengths and angles with critical biaxial strain applied are very closed to the ones under AC uniaxial strain. Furthermore, the phonon dispersions, phonon instabilities, and failure mechanism of these materials under three types of strains are also calculated and investigated.

## Computational details

All first-principles calculations are performed by using the Vienna ab initio simulation package (VASP)[32-34] based on the density functional theory (DFT). The generalized gradient approximations as parameterized by Perdew, Burke and Ernzerhof (PBE) is chosen for exchange-correlation functions.[35] The energy convergence criterion is chosen to be $10^{-8}$ eV. A Monkhorst-Pack[36] k-mesh of $15 \times 15 \times 1$ is used to sample the Brillouin Zone (BZ). The vacuum space of at least 20 Å is kept along the $z$ direction, which is enough to avoid the interactions between periodical images. All geometries are fully optimized until the maximal Hellmann-Feynman force is no larger than $10^{-4}$ eV Å$^{-1}$. And the cutoff energy of the plane-wave is 600 eV. The DFT-DF method is used to account for the long-range van der Waals interactions.[37,38] The phonon dispersions are obtained from the finite displacement method as implemented in the PHONOPY package.[39] A 6 × 6 × 1 supercell with a 3 × 3 × 1 k-mesh for the Brillouin zone sampling is used to ensure the convergence.

The primitive cell of these 2D materials has 2 atoms as shown in Figure 1a. However, we use an orthorhombic cell containing 4 atoms as displayed in Figure 1c, for applying uniaxial strains.[5] In order to obtain the stress-strain relation under uniaxial tensile strain applying along armchair (AC) / zigzag (ZZ) direction, a series of incremental tensile strains are applied along this direction, and the lattice along ZZ/AC is relaxed until the corresponding conjugate stress components is less than 0.01 GPa. For biaxial stress-strain calculations, a series of incremental equibiaxial tensile tensions are also applied and the atoms in the unite cell are fully relaxed. Note that under equibiaxial strain, 2D material with honeycomb structure is found undergoes a phase transition in which an optical phonon instability occurs at the K point of the Brillouin zone (BZ) resulting in a reduction of strength.[40,41] As this soft mode can only be captured in unit cells with hexagonal rings, a conventional cell which contains 6 atoms for each material (Figure 1d) is chosen while biaxial strain applied. The engineering tensile strain is defined as $\epsilon = (L - L_0)/L_0$, where $L$ is the strained lattice constants and $L_0$ is the original lattice constants, respectively. The 2D stress with a unit of N m$^{-1}$ is used to represent the strength of these 2D materials, which can be expressed by multiplying the Cauchy stresses and the thickness of the unit cell.

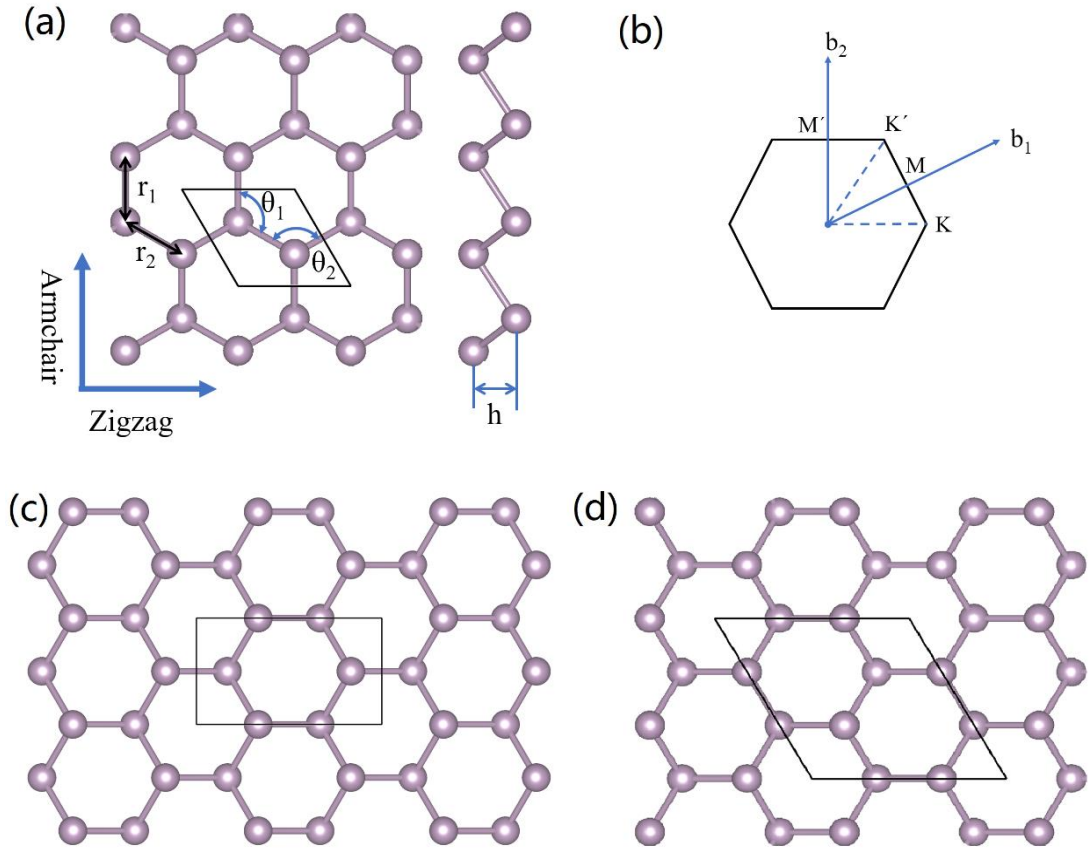

Figure 1. (a) The top view and side view of buckled hexagonal structure of monatomic group-V monolayer. The primitive cell containing 2 atoms is marked with solid lines. Note under uniaxial tensile strain, the bond lengths $r_1$ and $r_2$ are not equal, neither the bond angles $\theta_1$ and $\theta_2$. (b) shows the first Brillouin zone. It should be noted M´ and K´ become no longer equivalent to M and K under uniaxial tensile strain, as the symmetry of structure is broken. (c) and (d) show the unit cells containing 4 and 6 atoms, for applying uniaxial and biaxial strain, respectively. And the unit cells are displayed by solid lines.

## RESULTS AND DISCUSSION

We first optimized these 2D materials to determine the equilibrium lattice constants of them. It can be found all of them possess low-buckled configurations, which are different from the planar geometry of graphene. The optimized lattice constants and buckling heights are listed in Table 1, which are in good agreement with the previous work.[16,18,20]

Table 1. Calculated lattice parameters of blue phosphorene, arsenene, and antimonene, including lattice constant $a$, buckling height $h$, nearest-neighbor bond distance $r$ and angle $\theta$ between neighboring bonds.

|  | $a$(Å) | $h$(Å) | $r$(Å) | $\theta$(°) |
|---|---|---|---|---|
| Blue phosphorene | 3.268 | 1.245 | 2.261 | 92.57 |
| Arsenene | 3.585 | 1.409 | 2.503 | 91.43 |
| Antimonene | 4.077 | 1.662 | 2.881 | 90.04 |

The calculated stress-strain relations of phosphorene, arsenene, and antimonene under strains are presented in Figure 2. It can be found the stresses are proportional to strains in a very small range up to about 0.02 for the three 2D materials under uniaxial strains. In the small range, the near degeneracy of the elastic responses along AC and ZZ directions implies the in-plane elastic isotropy of these 2D materials, which results from the symmetry of hexagonal structure. As uniaxial strains increasing, the stress-strain relations become nonlinear, and we can find that the stresses caused by the uniaxial strains along AC direction increase faster than the ones along ZZ direction. It can be argued the in-plane symmetry is broken as materials deviate from the hexagonal structure dramatically while the uniaxial strains increasing. In contrast, the stress-strain relations under biaxial strains always show identity along AC and ZZ directions, as the symmetry of hexagonal structure is kept under biaxial strain.

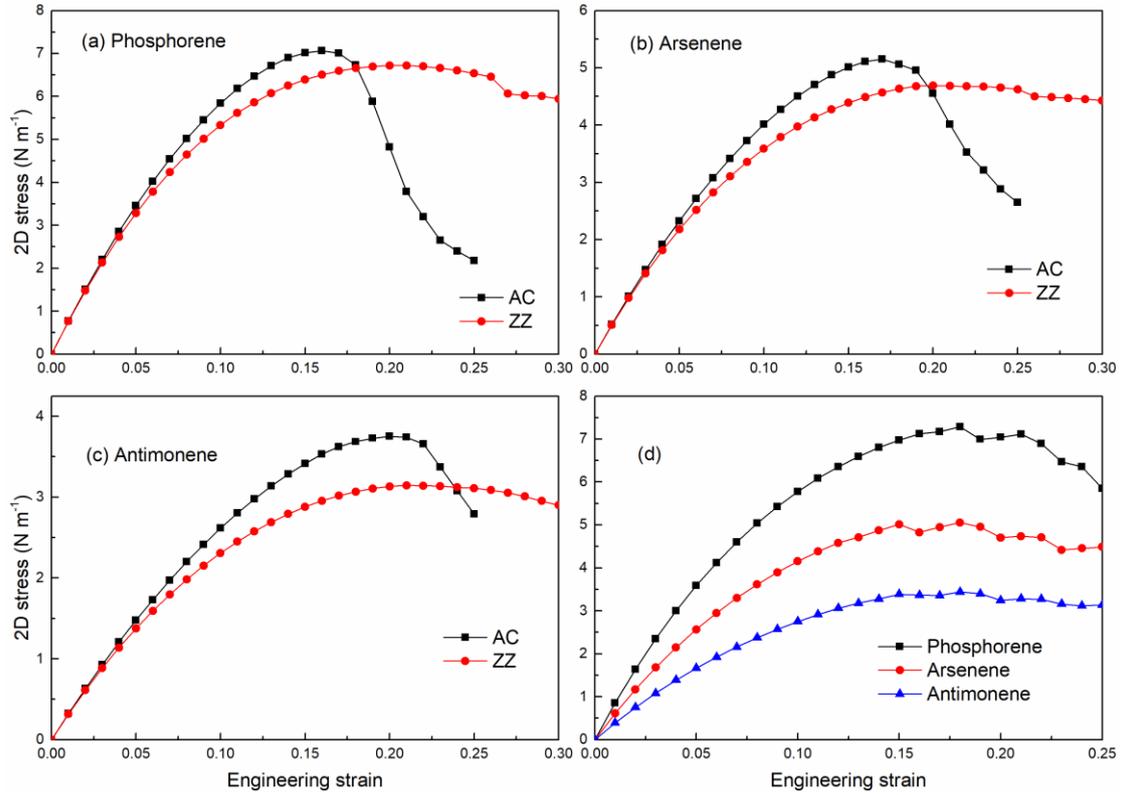

Figure 2. Calculated stress-strain relations for three 2D materials under different types of strains. Note that the stresses along AC and ZZ direction are always identical under biaxial strains, so there is only one curve for each 2D material under biaxial strains in (d).

The ideal strengths and corresponding critical strains for three materials are listed in Table 2. We find ideal strengths are in the range of about 3.7 to 7.0 N m$^{-1}$. Compared with other 2D materials, they are much lower than that of other 2D materials such as graphene,[40] graphane,[42] MoS$_2$[43,44] and borophene.[45] The critical uniaxial strains for the three 2D materials are in the range of 0.16 to 0.21. Under uniaxial strains, all the three 2D materials can sustain more deformation along ZZ direction than AC direction, while the ideal strengths along this direction are lower than those along AC direction. And the ideal strengths under biaxial strains are closed to the ones under those under uniaxial AC strains. It shows the chirality effects from the directional dependence of the critical strains. Typically, a hexagonal lattice structure can withstand more loading along ZZ direction than AC direction, such as graphene,[29,46] stanene,[47,48] and silicane.[49] Furthermore, from phosphorene to antimonene, we find their ideal strengths decrease. For instance, the ideal strength for phosphorene is about two times of that for antimonene along AC and ZZ directions. It can be attributed to the weaker atomic interaction as increasing atomic radius and bond length going down the Group-V atoms. It also can be found that the critical strain changes not so much from phosphorene to antimonene, especially the ones along ZZ direction, which are nearly equal to each for three materials. As everyone knows, there are various stable allotropes for Group-V 2D materials, and their critical strains show great differences from the ones we studied here. For instance, there are blue, black, green and red phosphorenes as the allotropes of 2D phosphorus [50-52] The critical strains of black phosphorene are 27% and 30% in the zigzag and armchair directions, respectively.[53] Moreover, green phosphorene can sustain uniaxial tensile strain up to 35% along the armchair direction.[54] For comparison, the blue phosphorene can sustain uniaxial tensile of 21% only along zigzag direction. It suggests that structure affects its critical strain significantly, and puckered structure can result in much better flexibility than buckled structure.

The 2D Young's modulus of three 2D materials are calculated. It is found the 2D Young's modulus are 79.7, 52.8 and 32.9 N m$^{-1}$ for phosphorene, arsenene, and antimonene. They are much smaller than borophene (389 and 166 N m$^{-1}$ along AC and ZZ direction),[45] but can be comparable to silicene (59.0~61.7 N m$^{-1}$),[55] germanene (43.4~44.0 N m$^{-1}$) [55] and stanene (26.7, 25.2 N m$^{-1}$).[47,55] It's well-known theoretically there is an upper limit of ideal strength for a material. It was firstly obtained by Griffith using extrapolation from experiment, and the value is about $E/9$ where $E$ is the Young's modulus.[31,47] Then subsequent researches determine it is of range $E/10 \sim E/9$.[56-59] Most bulk materials have an observable (realistic) strength several orders lower than this limit as there are lots of flaws and impurities which can decrease strength greatly. However, it is found that the ideal strength of many 2D materials approach the limit due to the quantum confinement effect. By experiments or density functional theory (DFT) simulations, it is demonstrated that the ideal strength of some 2D materials is in the range of $E/10 \sim E/15$, such as monolayer MoS$_2$, borophene, hexagonal boron nitride, and silicene.[45,60-62] And it is found theoretically that stanene possesses ideal strength of ~$E/7$ beyond the theoretical limit.[47] It is important to note the ideal strength of antimonene along AC direction breaks through the Griffith limit of $E/9$. In fact, our result suggests that the ratio of the Young's modulus to ideal strength is 8.75 for antimonene along AC direction. Moreover, the ratio becomes smaller from phosphorene to antimonene, and at last breaks through the conventional estimates of Griffith limit for antimonene.

Table 2. The ideal strengths (in N m$^{-1}$) and corresponding critical strains under different strains applied. The 2D Young's modulus $E$ (in N m$^{-1}$) and corresponding theoretical strength estimate, $E/10 – E/9$ are also listed.

|  | $\sigma_{AC}$ | $\epsilon_{AC}$ | $\sigma_{ZZ}$ | $\epsilon_{ZZ}$ | $\sigma_{bi}$ | $\epsilon_{bi}$ | $E$ | $E/10 – E/9$ |
|---|---|---|---|---|---|---|---|---|
| Phosphorene | 7.06 | 0.16 | 6.72 | 0.21 | 7.29 | 0.18 | 79.7 | 7.97 – 8.85 |
| Arsenene | 5.15 | 0.17 | 4.69 | 0.20 | 5.02 | 0.15 | 52.8 | 5.28 – 5.87 |
| Antimonene | **3.75** | 0.20 | 3.14 | 0.21 | 3.38 | 0.15 | 32.9 | 3.28 – 3.64 |

To further analyze the structural changes of the 2D materials under the three types of strains, in Figure 3 we show the ratio of strained bond length relative to that in unstrained material, $r/r_0$, where $r$ is the length of X–X (here X denotes P, As, and Sb atom respectively) bond in the structure under strain, while $r_0$ means the one in unstrained structure. In Figure 3, it shows that the strain dependent bond ratios are similar for the three materials. Under the uniaxial tension along AC direction, the bond $r_1$ increases monotonously with increasing strain, while $r_2$ is stretched slightly at first and then compressed. The inflexion points are about $\epsilon$ = 0.12, 0.13 and 0.17 for phosphorene, arsenene, and antimonene, respectively. With the increasing uniaxial strain applied along ZZ direction, the $r_1$ bond shows a nearly linear decrease, while the $r_2$ bond displays a linear increase. It is obvious that $r_1$ is stretched much more significantly under AC tensile strain than $r_2$ under ZZ strain. The strain dependence of bond length under biaxial strain which is not so large is very similar to the bond $r_1$ under uniaxial strain along AC direction.

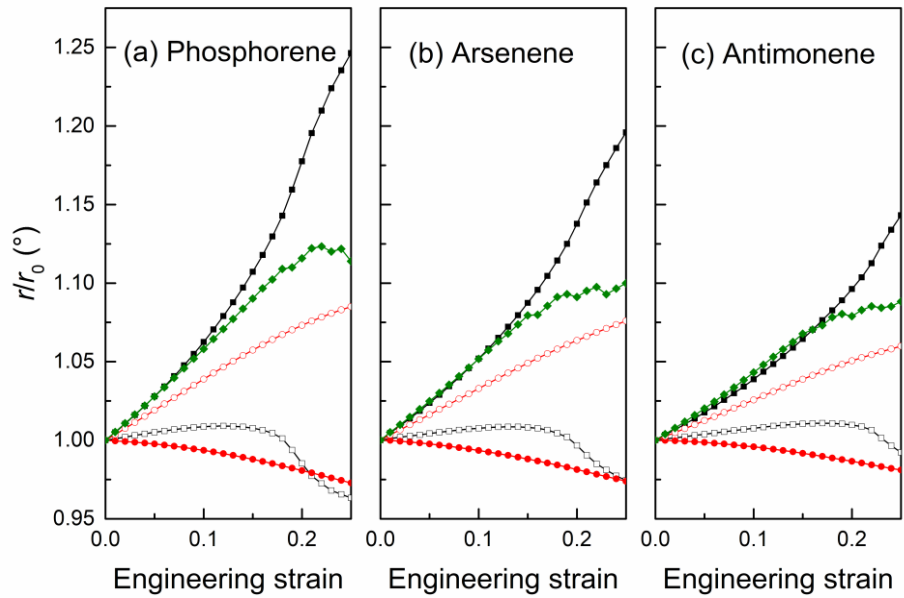

Figure 3. The strain dependent bond length for phosphorene, arsenene, and antimonene. Here black and red lines mean the ratios under uniaxial tension along AC and ZZ direction, while green lines represent the ratios under biaxial strain. The solid symbols represent $r_1/r_0$, and open ones express $r_2/r_0$. Note under biaxial strain bond $r_1$ is identical with $r_2$, so there is only one curve for each material.

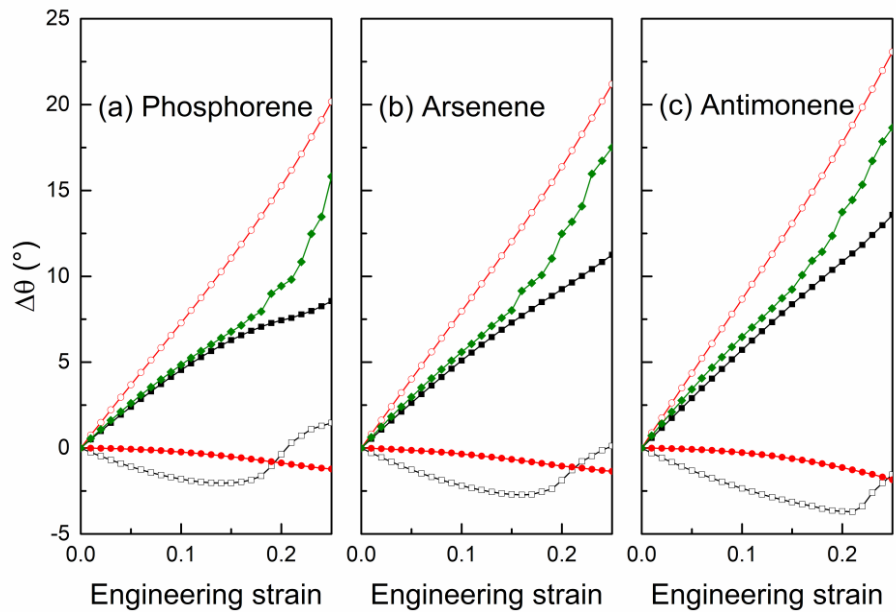

Figure. 4. The strain dependence of bond angles for three materials. The black and red lines mean the angle under uniaxial tension along AC and ZZ direction respectively, while green lines represent the one under biaxial strain. The solid symbols represent $\theta_1$, while open ones mean $\theta_2$. However, there is only one curve for each material under biaxial strain as $\theta_1$ is equal to $\theta_2$.

The variations of bond angles ($\Delta\theta_1$ and $\Delta\theta_2$) for three materials under different strains are investigated and plotted in Figure 4. Under the same type of strain, the curves for each material are similar. $\Delta\theta_1$ under uniaxial tension along AC direction increase monotonically with increasing strain, while $\Delta\theta_2$ decrease slowly at first and then rise. The inflection points are about 0.14, 0.16 and 0.21 for phosphorene, arsenene, and antimonene respectively. Note they are very close to their critical strains along AC direction. Under increasing ZZ strain, the curves of $\Delta\theta_2$ rise very fast in the whole range we studied, while the curves of $\Delta\theta_1$ decline very slowly, especially when the strains are small. Comparing the bond length under strains, it is found that AC strain stretch bond length $r_1$ most as $r_1$ is parallel to the direction of strain, while the bond angle changes not so much. Under ZZ strain, the bond angles $\theta_2$ change most but the bond lengths don't stretch much. That is to say, when bond length is stretched more, the variation of bond angle is less, and vice versa. Under biaxial tensile strain, we find the change of bond angles is very closed to that under AC uniaxial strain when the biaxial strain is less than critical strain.

Table 3. The stretched ratios of bond lengths and variations of bond angles under three different types of critical tensile strains for phosphorene, arsenene, and antimonene.

|  | AC | | | | ZZ | | | | Biaxial | |
| --- | --- | --- | --- | --- | --- | --- | --- | --- | --- | --- |
|  | $r_1/r_0$ | $r_2/r_0$ | $\Delta\theta_1$ | $\Delta\theta_2$ | $r_1/r_0$ | $r_2/r_0$ | $\Delta\theta_1$ | $\Delta\theta_2$ | $r/r_0$ | $\Delta\theta$ |
| Phosphorene | 1.130 | 1.005 | 6.8 | -1.9 | 0.979 | 1.076 | -1.0 | 16.2 | 1.109 | 7.9 |
| Arsenene | 1.105 | 1.007 | 8.1 | -2.7 | 0.981 | 1.063 | -1.1 | 16.4 | 1.079 | 8.0 |
| Antimonene | 1.096 | 1.010 | 10.9 | -3.7 | 0.985 | 1.053 | -1.3 | 18.8 | 1.068 | 9.2 |

Here we list in Table 3 the changes of bond lengths and angles for phosphorene, arsenene, and antimonene under critical strains. Under critical AC strain, bond $r_1$ is stretched seriously and $r_2$ changes much less. Meanwhile, bond angle $\theta_1$ changes much more than $\theta_2$. It can be attributed that $r_1$ is along the AC direction. Under AC tensile strain, the bond length of $r_2$ and the bond angle between $r_2$ are rigid and change little. Under the ZZ strain, $r_2$ varies much more than $r_1$, but is stretched less than $r_1$ under AC strain, as there is no bond parallel to ZZ direction. The bond angle $\theta_1$ changes little and $\theta_2$ changes a lot even much more than $\theta_1$ under AC strain. It can be argued when strain is applied along a bond, the stretch of the bond length is most significant, however the rotation of a bond is remarkable when the strain is inclined to it. Comprehensively considering the bond lengths and angles under different types of strains, it can be believed that the critical strain is determined by the stretch and rotation of bonds both. Under each type of tensile strain, the more the bond is stretched, the less it rotates, and vice versa.

In Figure 3 and Figure 4, it is obvious that the curves of strain dependent bond length and angle for biaxial strain lower than critical point are similar to the ones for AC strain. Moreover, as listed in Table 3, the bond lengths of critical biaxial strain are a little less than $r_1$ under AC uniaxial strain, while the critical bond angles are very closed to than $\theta_1$ under AC uniaxial strain, too. It can be attributed to that biaxial strain is composed of two uniaxial strains which are along two crystallographic axes. Since AC uniaxial strain is along one crystallographic axe, the bond length and angle with critical biaxial strain applied are very closed to the ones under AC uniaxial strain. Furthermore, in Table 3 we can also find that from phosphorene to antimonene, the stretched ratios $r_1/r_0$ and $r_2/r_0$ display inverse tendencies under critical AC and ZZ tensile strains, and they show the same tendencies under critical AC and biaxial strains. However, the variations of bond angles always get larger under all critical strains. It manifests that bonds are more likely to rotate under strains from phosphorene to antimonene.

Buckling height is a critical parameter for buckled 2D material. Different from the planar geometry of graphene, buckled configuration is found in the three group-V 2D materials. Since their long distance between atoms in the three 2D materials, the weak π-π orbital coupling between the nearest atoms make the planar structure unstable. They are stabilized by increasing overlap of $p_z$–$p_z$ through buckling, meanwhile the $sp^2$ hybrid orbitals are slightly dehybridized to form $sp^3$-like orbitals with $pz$.[63-66] Here we have calculated varied buckling heights under three types of strains for each material and show the ratios of buckling height $h$ for strained material to $h_0$ for strain-free structure in Figure 5. We can find the variations of buckling heights under uniaxial strains decrease monotonously, and are almost identical along AC and ZZ direction, especially when the strains lower than critical strain. And buckling heights decrease more quickly under biaxial strain than uniaxial strain. It is noteworthy that the buckling heights drop sharply beyond the critical biaxial strain, owing to the fact that material undergoes a structural transformation beyond critical biaxial strain. The ratios of $h/h_0$ for three materials under critical strains are listed in Table 4. As these materials can sustain more strain along ZZ direction, the buckling heights under critical strain along AC direction are always higher than those under critical strain along ZZ direction, except antimonenen since its critical strains are nearly the same along AC and ZZ directions. Furthermore, from phosphorene to antimonene, the variations of buckling heights at critical strain get larger. In contrast, it can be found that under critical AC, ZZ and biaxial strains, the changes of bond lengths and bond angles display different tendencies from phosphorene to antimonene. It is difficult to sum up a clear law based on the changes of bond length only since the effects of strain both changes of bond lengths and bond angles. However, buckling height $h$ is the vertical projection which can represent the combined effects of changes of bond lengths and bond angles. Buckling heights of these 2D materials get larger from phosphorene to antimonene with increasing bond length and weaker atomic interactions. They keep stable through larger buckling height which can result in more overlap of $p_z$–$p_z$ orbital couplings.[67] It can be argued from phosphorene to antimonene, the buckling height can reduce more under critical strains as it still can keep $p_z$–$p_z$ overlap enough to stabilize the structures. The overlap of $p_z$–$p_z$ plays an important role when strain applied.

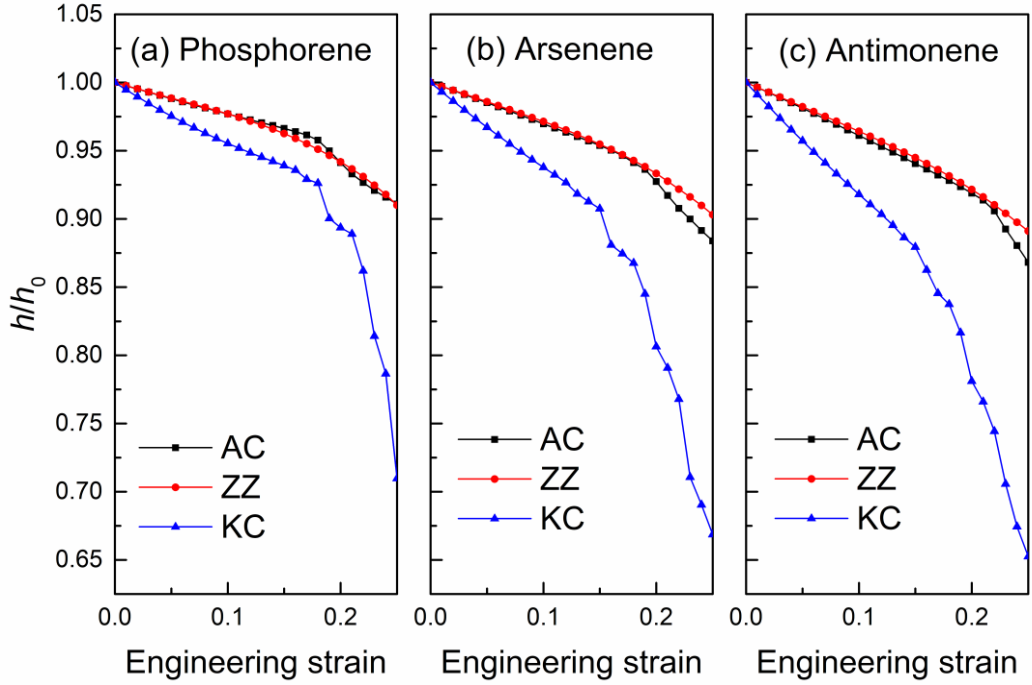

Figure 5. The ratios of buckling height *h* for strained structure to $h_0$ for strain-free structure.

Table 4. The ratios of buckling heights under three different types of critical tensile strains for phosphorene, arsenene, and antimonene.

|  | AC | ZZ | Biaxial |
|---|---|---|---|
| Phosphorene | 0.96 | 0.94 | 0.93 |
| Arsenene | 0.95 | 0.93 | 0.91 |
| Antimonene | 0.92 | 0.92 | 0.88 |

The trend of the evolution of the electronic charge density for the three monolayers under three type of critical strains is investigated and displayed in Figure 6. The case of the undeformed state is also exhibited in Figure 6a for comparison. Note we only show the charge density distributions of blue phosphorene, as those of other two 2D materials are very similar. The color of the bonding region qualitatively describes the extent of cohesion. In Figure 6b, it can be seen clearly that the electrons accumulated in the region between atoms along AC direction become fewer under critical AC strain. It indicates the weakening of the bond along AC direction and finally a separation of adjacent atoms along this direction. In Figure 6c and 6d, similar phenomena can be observed, implying separations of bonds corresponding to different critical strains.

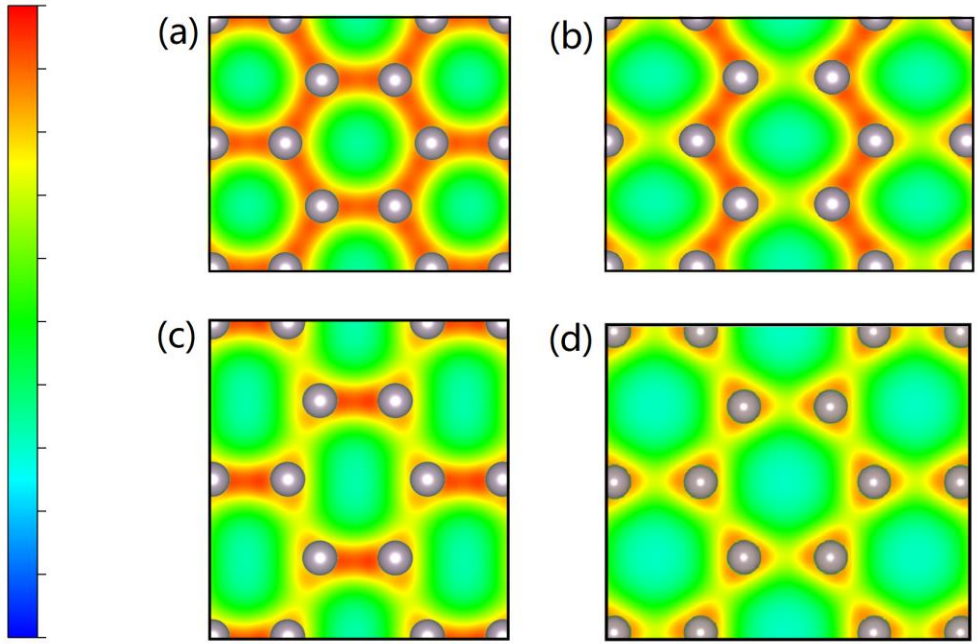

Figure 6. The electronic charge density distributions of blue phosphorene under different critical strains. (a) shows the case of unstrained structure. (b) and (c) display the distributions of phosphorene under critical AC and ZZ strains, while (d) shows the case of the structure under critical biaxial strain. Increasing electronic charge density from 0 to 0.13 is plotted with colours from navy to red.

Furthermore, we have calculated the electron localization function (ELF) of them and display here to analyze the evolution of bonding nature under different strains, as shown in Figure 7. Only the ELF of blue phosphorene is displayed here, as those of other two 2D materials are similar. In Figure 7(a), it can be found the regions of high value of ELF up or under phosphorus atoms, indicating the formation of π bond and overlapping of $p_z$-$p_z$ orbitals to maintain the stability of structure.[67] Additionally, the σ bond are also exhibited by high value regions between phosphorus atoms.[67] Under such strains, the regions of σ bond expand along the plane with the decrease of buckling heights, implying the increase of σ bond.

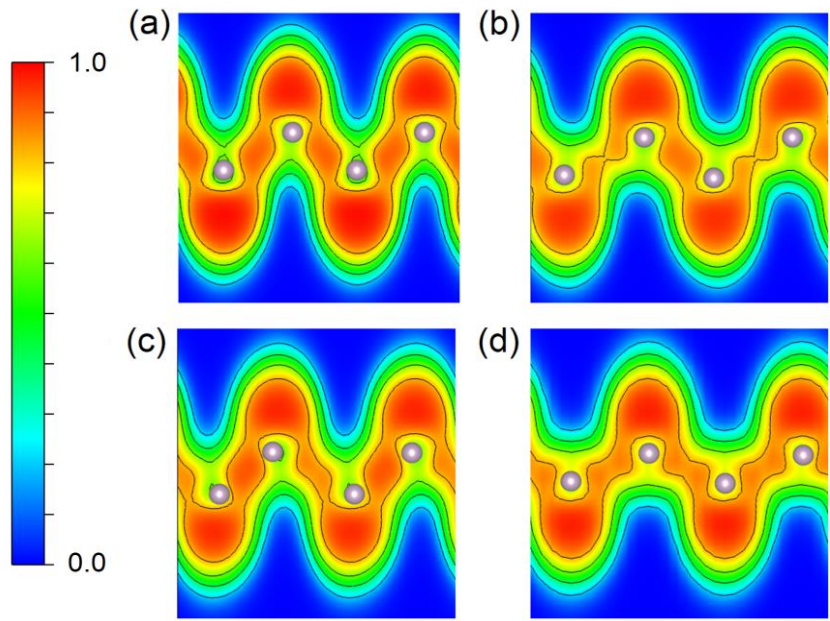

Figure 7. The electron localization functions of blue phosphorene under critical uniaxial strains along (b) AC and (c) ZZ directions. (d) shows the case under critical biaxial strain, and (a) is for the one free of any strain.

It is well known that the stability of materials is related to their phonons inherently. If there are imaginary frequencies along any high-symmetry direction of the Brillouin zone, it indicates the dynamical instability of the crystal structure. Here we calculated all the phonon dispersions of these 2D materials under different critical tensile strains and show them in Figure 8, while the phonon dispersions of free structures are also displayed for comparison. And all the failure mechanism under different types of strains are listed in Table 5 for the three 2D materials. It is found that all the three 2D materials possess wide phonon gap, which are about 4.3, 2.9 and 2.4 THz for phosphorene, arsenene, and antimonene, respectively. When critical strains applied, all the phonon gaps decrease drastically, even disappear under some types of strains. It's important to note that for all the 2D materials, the phonon dispersion of materials under the critical strain along ZZ direction shows imaginary frequencies along Γ-M, while there isn't any imaginary frequency under other types of critical strain. After examining the eigenvectors of the unstable phonon modes, it is found that all the soft modes are ZA modes, which are acoustic branches vibrating along the out-of-plane direction. It is the same to the case of strained $MoS_2$[43] and borophene,[45] but different from strained graphene.[29] The emergence of imaginary frequency shows that the 2D material would become instable before reaching the critical strains. In this case, the failure mechanism under these tensile strains are phonon instability instead of elastic instability.

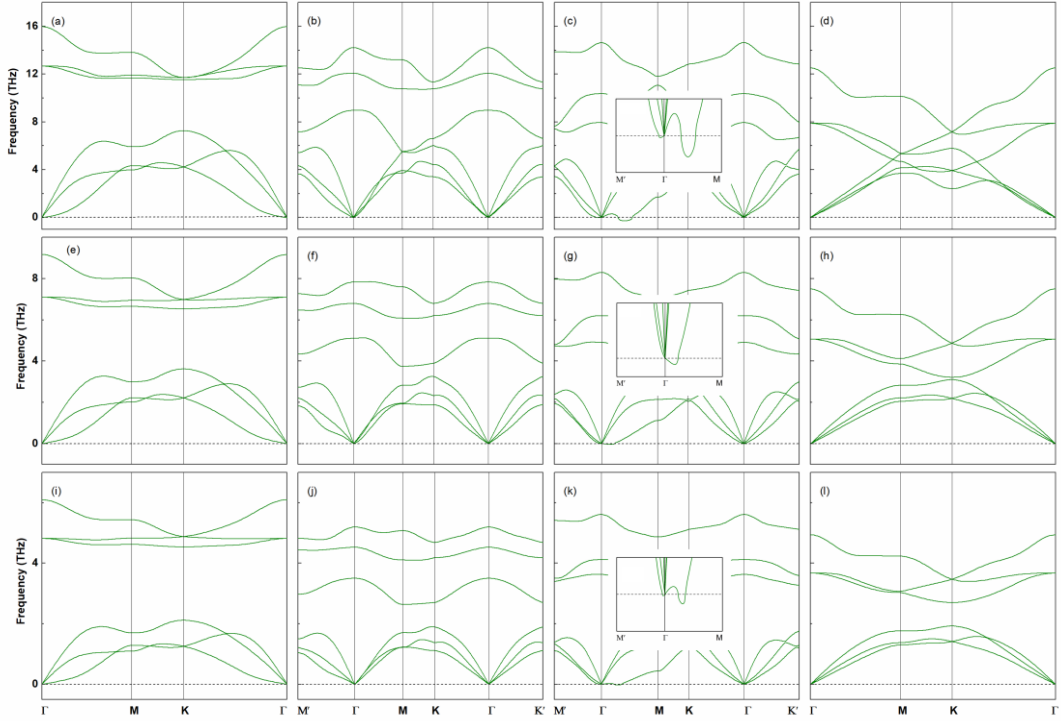

Figure 8. The phonon dispersions of blue phosphorene, arsenene, and antimonene. The top row shows the phonon dispersions of blue phosphorene, the middle row is for arsenene, and the bottom represents the phonon dispersions of antimonene. The four figures of each row display the phonons of structures under no strain, the critical AC strain, the critical ZZ strain, and the critical biaxial strain, respectively. The insets in (c), (g) and (k) show the details of negative frequencies.

Table 5. Failure mechanism for three 2D materials under different types of strains.

|  | AC | ZZ | Biaxial |
| --- | --- | --- | --- |
| Phosphorene | Elastic instability | Phonon instability | Elastic instability |
| Arsenene | Elastic instability | Phonon instability | Elastic instability |
| Antimonene | Elastic instability | Phonon instability | Elastic instability |

## Conclusion

In conclusion, the mechanical response of blue phosphorene, arsenene, and antimonene are investigated systematically based on first-principles calculations. The ideal strengths and critical strains for these 2D materials are studied under uniaxial and equibiaxial strains. It is found that the ideal strengths decrease greatly as the atomic number increases from phosphorene to antimonene. However, their critical strains change little under ZZ strain. Compared with their allotropes, it can be seen the critical strains are affect significantly by the structure. Interestingly, the ideal strength of antimonene along AC direction is found to exceed Griffith strength limit of $E/9$. Additionally, the bond lengths and bond angles are calculated under different types of strains. It is found the stretch of the bond length is most significant when the strain applied along a bond, while rotation of bond is remarkable when the strain is inclined to it. That is to say, the more the bond is stretched, the less it rotates, and vice versa. And it should be noted the changes of bond lengths and bond angles under biaxial strain are similar to those under AC strain. Furthermore, the buckling heights are also studied under different types of strains. It is found that the curves of buckling heights decline under increasing strains for each material, and they are almost identical under AC and ZZ strains. And the variational ratios of buckling heights under critical strain decrease from phosphorene to antimonene. The distributions of charge density are also investigated. The weakening of bonds and separations are observed under different critical strains. Phonon dispersions, phonon instabilities, and failure mechanism of three 2D materials under three types of strains are investigated and explained. The phonon instabilities occur for all the three 2D materials under critical ZZ strain. It is also found that all the soft modes are ZA branches.

## Acknowledgements

This work is supported by the National Key R&D Program of China (2016YFA0201104) and the National Science Foundation of China (Nos. 91622122 and 11474150). The use of the computational resources in the High Performance Computing Center of Nanjing University for this work is also acknowledged. Z.G. gratefully acknowledges the China Scholarship Council (CSC) for financial support (to C.S., 201706260027).

TOC Graphic

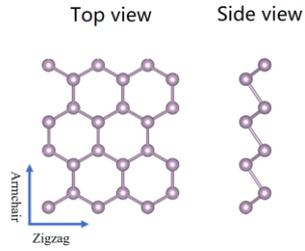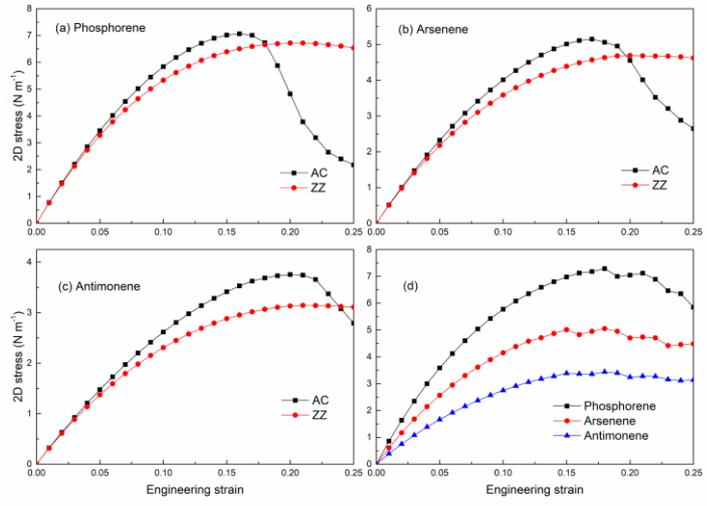